\documentclass[prl,floatfix,twocolumn,showpacs,preprintnumbers,amsmath,amssymb,superscriptaddress]{revtex4}
\usepackage{graphicx,psfrag} 
\usepackage{hyperref}
\newcommand{\newc}{\newcommand}
\newc{\gsim}{\lower.7ex\hbox{$\;\stackrel{\textstyle>}{\sim}\;$}}
\newc{\lsim}{\lower.7ex\hbox{$\;\stackrel{\textstyle<}{\sim}\;$}}
\newc{\gev}{\,{\rm GeV}}
\newc{\mev}{\,{\rm MeV}}
\newc{\ev}{\,{\rm eV}}
\newc{\kev}{\,{\rm keV}}
\newc{\tev}{\,{\rm TeV}}
\def\ln{\mathop{\rm ln}}

\newc{\mz}{M_Z}
\newc{\mpl}{M_*}
\newc{\mw}{m_{\rm weak}}
\newc{\nr}[1]{N^c_R{}_{#1}}
\usepackage{amsmath}
\def\beq{\begin{equation}}
\def\eeq{\end{equation}}
\def\bea{\begin{eqnarray}}
\def\eea{\end{eqnarray}}
\def\bitem{\begin{itemize}}
\def\eitem{\end{itemize}}
\newc{\ie}{{\it i.e.}}          \newc{\etal}{{\it et al.}}
\newc{\eg}{{\it e.g.}}          \newc{\etc}{{\it etc.}}
\newc{\cf}{{\it c.f.}}
\def\bar#1{\overline{#1}}
\def\vev#1{\left\langle #1 \right\rangle}

\def\inv{^{\raise.15ex\hbox{${\scriptscriptstyle -}$}\kern-.05em 1}}
\def\lbar{{\lower.35ex\hbox{$\mathchar'26$}\mkern-10mu\lambda}} %lambda bar

\def\to{\rightarrow}

\let\<=\langle
\let\>=\rangle

\let\+=\uparrow

\begin{document}   
\title{Maximally Natural Supersymmetry} 
%\date{\today}
\author{Savas Dimopoulos}
\email{savas@stanford.edu}
\affiliation{Department of Physics, Stanford University, Stanford, CA 94305, USA}
\author{Kiel Howe}
\email{howek@stanford.edu}
\affiliation{Department of Physics, Stanford University, Stanford, CA 94305, USA}
\affiliation{SLAC National Accelerator Laboratory, Menlo Park, CA  94025, USA}
\author{John March-Russell}  
\email{jmr@thphys.ox.ac.uk}  
\affiliation{Rudolf Peierls Centre for Theoretical Physics, University of Oxford,
1 Keble Rd., Oxford OX1 3NP, UK}
\affiliation{Department of Physics, Stanford University, Stanford, CA 94305, USA}
\begin{abstract}
We consider 4D weak scale theories arising from 5D supersymmetric (SUSY) theories with maximal Scherk-Schwarz breaking 
at a Kaluza-Klein (KK) scale of several \tev.  Many of the problems of conventional SUSY are avoided.  Apart from 3rd family sfermions the SUSY spectrum is heavy, with only $\sim 50\%$ tuning at a
gluino mass of $\sim 2 \tev$ and a stop mass of $\sim650 \gev$. A single Higgs doublet acquires a vacuum expectation value, so the physical Higgs is automatically Standard-Model-like.  A new $U(1)'$ interaction raises $m_h$ to $126\gev$. 
For minimal tuning the associated $Z'$, as well as the 3rd family sfermions, must be accessible to LHC13.
A gravitational wave signal consistent with BICEP2 is possible if inflation occurs when the extra dimensions are small.
\end{abstract}   
%\keywords{XXX}
%\preprint{}
\maketitle 
%%%%%%%%%%%%%%%%%%%%%%%%%%%%%%%%%%%%%%%%%%%%%%%%%%%%%%%%%%%%%%%%%%%%%%%%%%%%  
%\section{\label{intro}Introduction}
%%%%%%%%%%%%%%%%%%%%%%%%%%%%%%%%%%%%%%%%%%%%%%%%%%%%%%%%%%%%%%%%%%%%%%%%%%%%

%Weak-scale SUSY is in many ways the most attractive solution to the hierarchy problem, especially since the discovery of the Higgs scalar boson. 
The LHC has set stringent limits on the masses of SUSY particles and deviations in Higgs properties, implying a tuning of electroweak symmetry breaking (EWSB) at the percent level or worse for traditional SUSY models \cite{Gherghetta:2012gb,Arvanitaki:2013yja,Hardy:2013ywa,Feng:2013pwa,Gherghetta:2014xea,Fan:2014txa}.  This undermines the motivation for SUSY as the solution to the hierarchy problem and the case for discovery of SUSY at the LHC or proposed future colliders. 
% Although
%one can imagine that this tuning is a consequence of a relation between the SUSY parameters, in practice this is difficult to achieve in conventional models.
%The $\mu$ and $B_\mu$ parameters of the SUSY higgs sector presents a particular problem in this regard as $\mu$ directly contributes to the tuning of EWSB.
% while mechanisms generating $\mu$ and $B_\mu$ consistent with flavor constraints in general imply that there is no precise relation between the $\mu$ term and the sfermion mass terms $m_0$.   
%Another indication of this is that $\tan{\beta}$ is not determined in most SUSY theories, which implies that the top-Yukawa $\lambda_t$ is not determined.  Since $\lambda_t$ enters in the determination of $m_Z$, again the parameters have to be percent-tuned.   
Given the importance of this issue for current and future searches for new physics we examine the possibility of constructing natural, untuned theories.
%We find that this is the case.
Specifically, we consider 4D theories of the weak scale that arise from 5D SUSY theories with Scherk-Schwarz SUSY breaking (SSSB)
at a KK scale $1/R$ of several \tev  \cite{Antoniadis:1998sd,Delgado:1998qr,Pomarol:1998sd,Delgado:2001si,Delgado:2001xr,Barbieri:2003kn,Barbieri:2002sw,Barbieri:2002uk,Barbieri:2000vh,Marti:2002ar,Diego:2005mu,Diego:2006py,vonGersdorff:2007kk,Bhattacharyya:2012ct}. The key features are:

%\\
$\bullet$ The theory is never well approximated by a 4D softly-broken $N=1$ SUSY limit. Many of the problems
of the MSSM and its extensions are avoided.

%\\
$\bullet$ Higgsinos, gauginos, and the 1st and 2nd family sfermions get (mainly Dirac) SSSB masses of size $1/2R$.

%\\
$\bullet$ A natural SUSY spectrum \cite{Dimopoulos:1995mi,Pomarol:1995xc,Cohen:1996vb} is obtained through localization of the 3rd family on a 4D brane. The absence of  large logs due to the super-softness of SSSB \cite{Antoniadis:1997zg,Barbieri:2001dm,Delgado:2001ex,Contino:2001gz,Weiner:2001ui,Kim:2001re,Puchwein:2003jq} protects the weak scale and suppresses the tendency of the gluino to pull up the stop mass \cite{Arvanitaki:2013yja,Hardy:2013ywa}. 
%Apart from the 3rd generation sfermions and one mainly singlino state the SUSY spectrum is remarkably heavy, with only $\sim 30\%$ tuning at a gluino mass of $\sim 2 \tev$.\\

%\\
$\bullet$ The $\mu$ term neither exists nor is needed. Only $H_u$ acquires a VEV, and
the down-like quark and lepton masses are generated by Kahler couplings to $H_u^{\dagger}$ \cite{Davies:2011mp}. The physical
Higgs is automatically SM-like.

%\\
$\bullet$ An additional SUSY breaking sector is necessarily present for radius stabilization with zero cosmological constant (CC), and SUSY breaking in this sector can naturally be driven by SSSB.  Higher dimensional couplings of the MSSM fields to this sector play a crucial role in EWSB and collider phenomenology. 

$\bullet$ A $U(1)'$ broken in this additional sector raises the Higgs mass to $126~\gev$ through an unusual non-decoupling D-term, with a $Z'$ mass of order $1/R$.

The pattern of localization of matter and Higgs multiplets and the mechanism driving EWSB, generating Yukawa couplings, and accommodating the observed physical Higgs mass lead to important differences from previously studied models of SSSB \cite{Antoniadis:1998sd,Delgado:1998qr,Pomarol:1998sd,Delgado:2001si,Delgado:2001xr,Barbieri:2003kn,Barbieri:2002sw,Barbieri:2002uk,Barbieri:2000vh,Marti:2002ar,Diego:2005mu,Diego:2006py,vonGersdorff:2007kk,Bhattacharyya:2012ct}.

%EWSB and higgs properties approximately realise the $\tan{\beta}\to\infty$ limit of the MSSM but without the problems
%usually associated to this limit.
%$\bullet$ All SUSY parameters will be calculable, finite, and UV-insensitive. In particular there are no large logs, as the ``messenger scale'' coincides with the
%SUSY-breaking scale and the KK-scale. 

%%%%%%%%%%%%%%%%%%%%%%%%%%%%%%%%%%%%%%%%%%%%%%%%%%%%%%%%%%%%%%%%%%%%%%%%%%%%  
%\section{\label{intro}Minimal Scherk-Schwarz Module}
\section{Natural spectrum from Scherk-Schwarz}
%%%%%%%%%%%%%%%%%%%%%%%%%%%%%%%%%%%%%%%%%%%%%%%%%%%%%%%%%%%%%%%%%%%%%%%%%%%%

%We now introduce the minimal 5D  ``module" that will underly all our constructions.   
Symmetries may be broken in a way preserving 4D Poincare invariance by imposing boundary conditions (bc's) on bulk
fields involving a symmetry twist.  If the twist includes an R-symmetry group action, then SUSY is softly broken by the SSSB mechanism \cite{Scherk:1978ta, Scherk:1979zr}.
This SSSB is non-local from the higher-dimensional perspective, and is of an exceptionally soft type, similar to finite-temperature breaking of SUSY. 
In our case the twists will be maximal, $\pm 1$, and the underlying
non-gravitational sector can be described as a 5D gauge theory compactified on a $S^1/(Z_2 \times Z'_2)$ orbifold. 
The 5th dimension, of physical length $\pi R$, is parameterised by $y\in [0,\pi]$, and branes sit at
the inequivalent fixed points at $0, \pi R$. 

Our 5D bulk theory is a SUSY theory
containing the SM gauge fields, the first two families, and a pair of distinct Higgs hypermultiplets, $H_u, H_d$,  (see Fig.1a).  
As the minimal SUSY in 5D 
corresponds to $N=2$ 4D SUSY, the superpartners of these bulk states fill out $N=2$ 4D multiplets, with each 5D vector implying both a 4D vector and chiral supermultiplet in the adjoint representation, $V^a_{5D} = \{ V^a_{4D}, \bar{\Sigma}^a \}$ (with physical fields $V^a_{\mu},\lambda^a$ and $\bar{\sigma}^a,\bar{\lambda}^a$) while the matter fields are hypermultiplets consisting of 4D chiral and anti-chiral multiplets $\Phi^i_{5D} = \{ \phi^i, \bar{\phi}^i \}$ (physical fields $\varphi_{i},\psi_{i}$ and $\bar{\varphi}_{i}, \bar{ \psi}_{i}$) \cite{Marcus:1983wb,ArkaniHamed:2001tb,Marti:2001iw,Hebecker:2001ke,Linch:2002wg}.

The two $Z_2$ actions, at their respective fixed points at $0,\pi R$, break 5D SUSY to two {\it different} and
incompatible $N=1$ 4D SUSYs thus breaking SUSY completely in the 4D effective theory; the component field bc's are summarised in Table~\ref{bcs}.  
Due to the non-local nature of SSSB there are {\it no cutoff-dependent log enhancements} of the effective 4D soft terms.  
At $y=0$ we localise the 3rd generation fields. As the fixed points preserve only $N=1$ 4D SUSY, these
states are simply 4D chiral multiplets with no additional partners, and a localised Yukawa superpotential for up-like states is allowed
\begin{gather}
\delta(y)  H_u(y) \biggl(\frac{\tilde{y_t}}{M_5^{1/2}} Q_3 U_3^c+\frac{\tilde{y_c}}{M_5^{3/2}}Q_2(y) U_2^c(y)+...\biggr).
\label{upyukawa}
\end{gather}
where $\tilde{y_i}$ are dimensionless and the Yukawa couplings to bulk 1st/2nd generations are naturally suppressed compared to the brane-localized 3rd generation. We later return to the down-type Yukawas.

\begin{table}\centering
\begin{tabular}{|c|c|c|c|c|}
%Row: 1
\cline{1-5}
\vbox to2.5ex{\vspace{1pt}\vfil\hbox to11ex{\hfil $$\hfil}} & 
\vbox to2.5ex{\vspace{1pt}\vfil\hbox to11ex{\hfil $(+,+)$ \hfil}} & 
\vbox to2.5ex{\vspace{1pt}\vfil\hbox to11ex{\hfil $(+,-)$\hfil}} & 
\vbox to2.5ex{\vspace{1pt}\vfil\hbox to11ex{\hfil $(-,+)$\hfil}} & 
\vbox to2.5ex{\vspace{1pt}\vfil\hbox to11ex{\hfil $(-,-)$\hfil}} \\

%Row: 2
\cline{1-5}
\vbox to2.5ex{\vspace{1pt}\vfil\hbox to11ex{\hfil $V^a_{5D}$\hfil}} & 
\vbox to2.5ex{\vspace{1pt}\vfil\hbox to11ex{\hfil $V^a_\mu$\hfil}} & 
\vbox to2.5ex{\vspace{1pt}\vfil\hbox to11ex{\hfil $\lambda^a $\hfil}} &  
\vbox to2.5ex{\vspace{1pt}\vfil\hbox to11ex{\hfil $\bar{\lambda}^a $\hfil}} & 
\vbox to2.5ex{\vspace{1pt}\vfil\hbox to11ex{\hfil $\bar{\sigma}^a $ \hfil}} \\

%Row: 3
\cline{1-5}
\vbox to2.5ex{\vspace{1pt}\vfil\hbox to11ex{\hfil $H_{u,d}$\hfil}} & 
\vbox to2.5ex{\vspace{1pt}\vfil\hbox to11ex{\hfil $ h_{u,d} $\hfil}} & 
\vbox to2.5ex{\vspace{1pt}\vfil\hbox to11ex{\hfil $ \psi_{h_{u,d}} $\hfil}} &  
\vbox to2.5ex{\vspace{1pt}\vfil\hbox to11ex{\hfil $ \bar{\psi}_{h_{u,d}}  $\hfil}} & 
\vbox to2.5ex{\vspace{1pt}\vfil\hbox to11ex{\hfil $ \bar{h}_{u,d} $ \hfil}} \\

%Row: 3
\cline{1-5}
\vbox to2.5ex{\vspace{1pt}\vfil\hbox to11ex{\hfil $F_{i=1,2}$\hfil}} & 
\vbox to2.5ex{\vspace{1pt}\vfil\hbox to11ex{\hfil $ \psi_{F_i} $\hfil}} & 
\vbox to2.5ex{\vspace{1pt}\vfil\hbox to11ex{\hfil $ \varphi_{F_i} $\hfil}} &  
\vbox to2.5ex{\vspace{1pt}\vfil\hbox to11ex{\hfil $ \bar{\varphi}_{F_i}  $\hfil}} & 
\vbox to2.5ex{\vspace{1pt}\vfil\hbox to11ex{\hfil $ \bar{ \psi}_{F_i} $ \hfil}} \\

%Row: 3
\cline{1-5}
\vbox to2.5ex{\vspace{1pt}\vfil\hbox to11ex{\hfil $ \Phi_{1,2}$\hfil}} & 
\vbox to2.5ex{\vspace{1pt}\vfil\hbox to11ex{\hfil $ \psi_{\Phi_{1,2}}  $\hfil}} & 
\vbox to2.5ex{\vspace{1pt}\vfil\hbox to11ex{\hfil $ \varphi_{1,2}$\hfil}} &  
\vbox to2.5ex{\vspace{1pt}\vfil\hbox to11ex{\hfil $ \bar{\varphi}_{1,2}$\hfil}} & 
\vbox to2.5ex{\vspace{1pt}\vfil\hbox to11ex{\hfil $ \bar{\psi}_{\Phi_{1,2}} $ \hfil}} \\

\cline{1-5}
\end{tabular}
\caption{Bc's at $y=(0,\pi)$ for bulk fields of complete model with $\pm$ corresponding to Neumann/Dirichlet.
%Rows show 5D N=1 supermultiplet content. Differing component field bc's non-locally (and fully) break SUSY. 
Only the $(+,+)$ fields have a zero mode, and the KK mass spectrum ($n \geqslant 0 $) is: $m_n=n/R$ for $(+,+)$ fields; $(2n+1)/2R$ for $(+,-)$ and $(-,+)$;  and $(n+1)/R$ for $(-,-)$. $\psi_{F_{1,2}}$ stands for all 1st/2nd
generation fermions; $\varphi_{F_i}$ their 4D $N=1$ sfermion partners; barred states are the extra 5D $N=1$ SUSY partners. 
%Zero modes along with brane localised 3rd family realise SM plus other light states. 
%At each $k/2R$ level states mix to produce mass eigenstates.
\label{bcs}}
\end{table}

There is no need for a $\mu$ term linking $H_u H_d$ to lift the higgsinos. Instead, SSSB gives the higgsinos a large $1/2R$ mass by marrying $ \psi_{h_u} $ with $\bar{\psi}_{h_u}$. The SSSB bc's lift the Higgsinos while making {\it no contribution} to the scalar Higgs masses, avoiding the usual source of tree-level tuning. 
%We will also show there is no need for a $H_d$ VEV. 

% By locating the first two families in the bulk, while the 3rd family is on a brane, the bulk states feel tree level SSSB with mass $1/2R$, while the 3rd family sfermions only get a mass at 1-loop, so we easily realize a natural SUSY scenario \cite{SDGian}.

After SSSB the brane-localised scalars pick up, at 1-loop, finite positive soft SUSY-breaking masses 
\begin{gather}
\delta{\tilde m_i}^2 \simeq  \frac{7 \zeta(3)}{16 \pi^4 R^2} \biggl( \sum_{I=1,2,3} C_I(i) g_I^2 + C_t(i) y_t^2 \biggr),
\label{softmasssq}
\end{gather}
with  $C(U_3) = \{ 4/9 , 0 , 4/3, 1 \}$, $C(D_3) = \{ 1/9 , 0 , 4/3, 0 \}$, $C(E_3) = \{ 1, 0, 0,0 \}$, $C(L_3) = \{ 1/4, 3/4, 0,0 \}$, $C(Q_3) = \{ 1/36, 3/4, 4/3,1/2 \}$, and
for the Higgs bulk scalar zero mode $C(H_{u,d}) = \{ 1/4, 3/4, 0,0 \}$ \cite{Antoniadis:1998sd}.

%%%%%%%%%%%%%%%%%%%%%%%%%%%%%%%%%%%%%%%%%%%%%%%%%%%%%%%%%%%%%%%%%%%%%%%%%%%
\begin{figure}
\newlength{\picwidtha}
\setlength{\picwidtha}{3.3in}
 \begin{flushleft}
{\resizebox{\picwidtha}{!}{\includegraphics{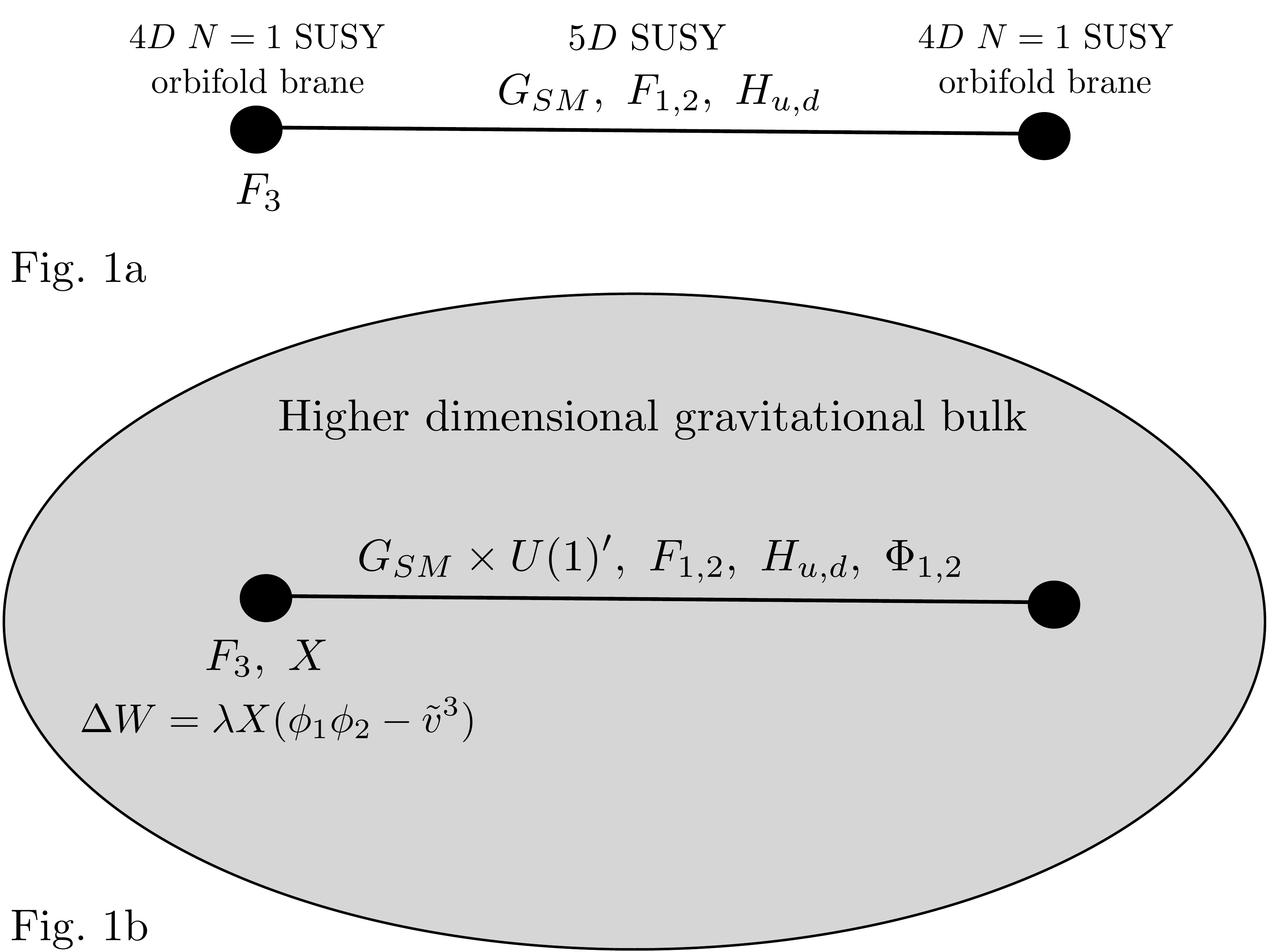}}}
 \end{flushleft}
\caption{\label{fig1} (a) Schematic of minimal model. In 5D are the SM gauge fields, the first two families $F_{1,2}$, Higgs doublets
$H_{u,d}$, and superpartners implied by 5D SUSY.  The 3rd generation chiral multiplets are brane-localised. SUSY is broken non-locally by bc's.
(b) Full model including embedding in yet higher-dimensional bulk.  The 5D $U(1)'$ is broken via $y$-dependent VEVs
(driven by the brane-localised superpotential $\Delta W$) of bulk fields, $\Phi_{1,2}$, of charges $\pm 1$. After SSSB, $F_X \sim 1/R^2$ is induced
for $X$, a brane-localised singlet field.} 
\end{figure}
%%%%%%%%%%%%%%%%%%%%%%%%%%%%%%%%%%%%%%%%%%%%%%%%%%%%%%%%%%%%%%%%%%%%%%%%%%%

%%%%%%%%%%%%%%%%%%%%%%%%%%%%%%%%%%%%%%%%%%%%%%%%%%%%%%%%%%%%%%%%%%%%%%%%%%%
\begin{figure}
%\newlength{\picwidtha}
\setlength{\picwidtha}{3.5in}
 \begin{flushleft}
{\resizebox{\picwidtha}{!}{\includegraphics{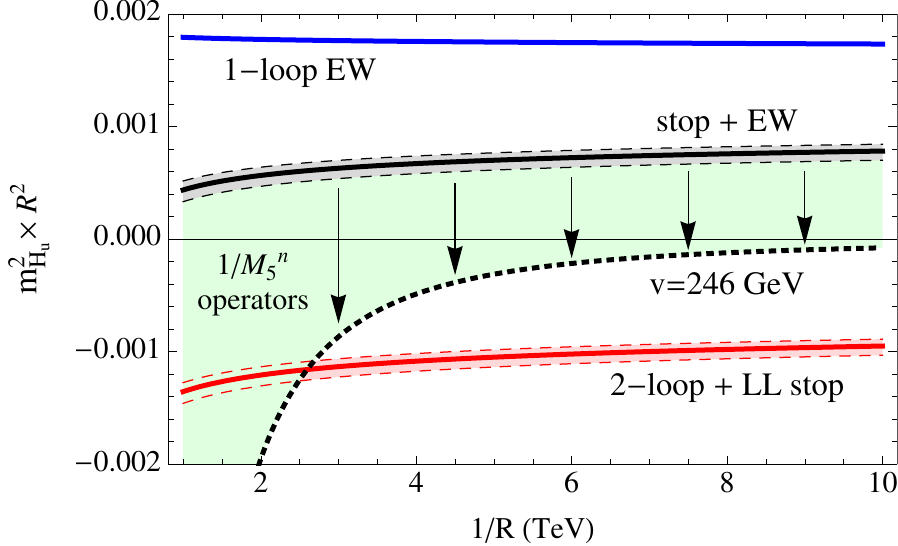}}}
 \end{flushleft}
\caption{\label{ewsbplot}Contributions to the Higgs soft mass $m^2_{H_u}$ in units of $1/R^2$. The positive 1-loop electroweak contribution (blue) and the negative 2-loop + leading log top-stop sector contribution (red) combine to give a positive mass squared (black). Contributions from higher-dimension operators Eq.(\ref{higgsHDO}) can lead to successful EWSB, indicated by the dotted black curve. The dashed bands show the uncertainty for $\bar{\rm MS}$ top mass $m_t(M_t) =160^{+5}_{-4} \gev$.} 
\end{figure}
%%%%%%%%%%%%%%%%%%%%%%%%%%%%%%%%%%%%%%%%%%%%%%%%%%%%%%%%%%%%%%%%%%%%%%%%%%%

In addition to  the positive 1-loop EW contribution Eq.(\ref{softmasssq}),
the Higgs soft mass $\tilde m_{H_u}^2$ receives a comparable  negative contribution at 2-loops from the $t$-${\tilde t}$ sector.  Ref.~\cite{Barbieri:2003kn} performed a 2-loop 5D calculation of this term, and we have used RG methods to  determine the leading 3-loop $\log(m_t R)$-,$\log(m_{\tilde t_1}/m_t)$-enhanced corrections, which are numerically important in determing the fate of EWSB \cite{longpaper}.  As shown in Fig.~\ref{ewsbplot}, these minimal contributions do {\it not}  so far lead to EWSB.  Nevertheless, the model has attractive features: Compared to 4D theories
the Higgs soft mass is more screened from SUSY-breaking as Eq.(\ref{softmasssq}) involves a finite 1-loop factor with no log
enhancement, SUSY breaking for all but the 3rd generation and Higgs scalar zero mode is {\it direct and universal},  and  higgsinos are heavy without a large $\mu$ term.

%%%%%%%%%%%%%%%%%%%%%%%%%%%%%%%%%%%%%%%%%%%%%%%%%%%%%%%%%%%%%%%%%%%%%%%%%%%
\section{\label{full}Successful EWSB and Higgs Mass}
%\section{\label{full}A Successful Un-tuned Theory}
%%%%%%%%%%%%%%%%%%%%%%%%%%%%%%%%%%%%%%%%%%%%%%%%%%%%%%%%%%%%%%%%%%%%%%%%%%%

Other faults remain in this model, and we find their solution
plays a major role for EWSB and experimental signatures.  
First, our 5D theory is an effective theory which must be cutoff at a scale $M_5$. The bulk 5D gauge couplings are dimensionful ($1/g_{I,4}^2 = \pi R/g_{I,5}^2$ up to small brane-kinetic-term corrections), and 5D perturbative unitarity bounds for $g_3$ require $\pi M_5 R \lesssim 25$ \cite{Muck:2004br,Chivukula:2003kq}. NDA strong coupling estimates for the brane-localized Yukawas give a similar bound \cite{Marandella:2004xm}.

This cutoff is large enough to justify the 5D viewpoint and the parameterization of UV effects in terms of higher dimensional operators, but the weakness of gravity in the low energy 4D theory, $M_{\rm pl} \gg M_5$, must still be explained. The two controllable possibilities of which we are aware are: (a) Embed the 5D theory in a 10 or 11D string/M-theory where some or all of the extra 5 or 6 {\it purely gravitational}
dimensions are `large', similar to the original brane-world proposal of Refs.\cite{ArkaniHamed:1998nn,Antoniadis:1998ig,ArkaniHamed:1998vp,Argyres:1998qn} (see Fig.1b).  Since our fundamental scale is $M_5 \gsim 30\tev$, $n\geq2$ extra dimensions is safe from cosmological, astrophysical, and laboratory constraints. (b) Utilise a little-string-theory construction with tiny string coupling \cite{Antoniadis:2011qw}.

Second, the radius $R$ is unstabilized. 
Moreover, SSSB without radius stabilization is of no-scale type with zero CC at tree level \cite{Luty:2002hj,Marti:2001iw,Kaplan:2001cg}, and generally radius stabilization yields a deep negative CC of order $\sim -1/R^4$ \cite{Ponton:2001hq,vonGersdorff:2003rq,Rattazzi:2003rj,vonGersdorff:2005ce,Dudas:2005vna}. An additional positive SUSY breaking sector (which may or may not be sequestered from the radion at tree level) can tune the minimum to zero CC, and will generally involve a brane-localised field, $X$, with an
F-term, $F_X\sim 1/R^2$ \cite{longpaper}.
%This dynamics can be induced by SSSB and so again has no log enhancements. 

With this additional brane-localized SUSY breaking $F_X\neq 0$, the Kahler operators 
\bea
\Delta {\cal K}_{m^2_{H}} &=& \delta(y) \frac{c_H}{M_5^3} X^\dagger X H_u^\dagger H_u\\
\Delta {\cal K}_{m^2_{\tilde{t}}} &=& \delta(y)   X^\dagger X \left( \frac{c_Q}{M_5^2} Q_3^\dagger Q_3 + \frac{c_U}{M_5^2} U_3^{c\dagger} U^c_3 \right)
\label{higgsHDO}
\eea
can alter the $H_u$ soft mass and trigger EWSB, either directly for $\Delta {\cal K}_{m^2_{H}}$ or radiatively through one-loop stop corrections for $\Delta {\cal K}_{m^2_{\tilde{t}}}$.  When the 5D picture is under good control, $(\pi R M_5) \gtrsim 10$, the contribution from $\Delta {\cal K}_{m^2_{\tilde{t}}}$ dominates.  As illustrated in Fig.~\ref{ewsbplot}, we find that for $F_X\sim 1/R^2$, and for $c_Q, c_U\sim 1$
this shift is sufficient to trigger EWSB at scales $1/2R \gsim 2\tev$. The tuning involved will be seen to be exceptionally mild for present collider limits. 

The bottom and tau Yukawas also result from $F_X$ via the Kahler terms~\cite{Davies:2011mp}
\begin{gather}
\delta(y)  (H_u(y)^\dagger X^\dagger) \biggl(\frac{\tilde{y_b}}{M_5^{5/2}} Q_3 D_3^c+...\biggr).
\label{downY}
\end{gather}
The 1st and 2nd generation down-type Yukawas can be generated by similar higher dimensional Kahler operators or by superpotential couplings to $h_u^\dagger$ on the $y=\pi$ brane \cite{Barbieri:2002uk}. Therefore $H_d$ need not obtain a VEV, a dramatic simplification of the Higgs sector only possible because the cutoff scale is so low. 
Although $H_d$ must be present to avoid a quadratically divergent Fayet-Illiopoulos (FI) term \cite{Ghilencea:2001bw,Marti:2002ar,Barbieri:2002ic}, unlike the MSSM there is no need for there to be $\mu$- or $B_\mu$-terms that link $H_u$ to $H_d$.  The simplest option is to impose an unbroken $Z_2$ symmetry on $H_d$ which forbids these
unnecessary terms and eliminates potentially dangerous flavor-changing effects;  $H_d$ is a stable (neutral) particle in the spectrum in addition to the LSP, as in the inert doublet
models \cite{Cirelli:2005uq,Barbieri:2006dq,Ma:2006km,LopezHonorez:2006gr}.

For $m_{\tilde{t}_1}\gtrsim 3.5\tev$, the radiative contributions to the physical Higgs mass may be large enough to accommodate $m_{h}=126\gev$  \cite{Feng:2013tvd}. For lighter stops, we obtain $m_{h}=126\gev$ with a non-decoupling D-term (as only $\vev{H_u} \neq 0$, a NMSSM-like singlet interaction $S H_u H_d$ can not be employed as in ref.~\cite{Delgado:2000fs}.).  Specifically, we introduce a bulk
$U(1)'$ gauging a subgroup of right-handed $SU(2)$ generated by $T_{3R}$ under which $H_u$ (and $H_d$) transform (the $U(1)'$ is anomaly-free if three light RHD neutrino superfields are introduced in the bulk; we find our theory allows a novel theory of neutrino mass generation~\cite{longpaper}). To avoid suppression of the quartic, the breaking of the new gauge group must couple to large SUSY breaking F-terms \cite{Batra:2003nj,Maloney:2004rc,Cheung:2012zq}. It is natural to associate the breaking of the $U(1)'$ with the same dynamics that generates $F_X$, with the resulting $Z'$ mass  $\sim 1/R$.

A simple model where $F_X$ is induced by SSSB and is associated with the breaking of the $U(1)'$ is obtained by introducing bulk hypermultiplet fields $\phi_{1,2}$ charged $\pm\frac{1}{2}$ under the $U(1)'$ with SSSB bc's  given in Table.~1 and a brane-localized superpotential
\begin{gather}
\Delta W=\frac{\lambda}{M_5} X\left( \phi_1(y)\phi_2(y)-\tilde{v}^3 \right) \delta(y).
\label{extraW}
\end{gather}
This leads to spontaneous breaking of the $U(1)'$ in the D-flat direction with a $y$-dependent profile for $\vev{\phi_{1,2}}$ and a brane-localized $F_X = M_5/(\lambda \pi R)$. This positive SUSY breaking contribution to the radion potential can be tuned to allow stabilization with zero CC. We find that for  $m_{\tilde{t}}\gtrsim650\gev~(m_{\tilde{t}}\gtrsim1\tev)$ and $m_Z' \lesssim 2/R$, the $U(1)'$ D-term can yield $m_{h}=126\gev$ with $g_X<1~(g_X<g_2)$.
The $U(1)'$ sector also contributes to the Higgs soft mass. 
The contribution is not well-approximated by the truncated lightest KK modes; we evaluate it in the 5D theory and find for $m_{Z'}\gtrsim 1/R$ the contribution favors EWSB and numerically approaches
\begin{gather}
\delta {m^2_{H_u}}(U(1)') \approx -10^{-3} g^2_{X} m^2_{Z'}.
\end{gather}

%%%%%%%%%%%%%%%%%%%%%%%%%%%%%%%%%%%%%%%%%%%%%%%%%%%%%%%%%%%%%%%%%%%%%%%%%%%
\section{\label{pheno} Phenomenology and Variations}
%%%%%%%%%%%%%%%%%%%%%%%%%%%%%%%%%%%%%%%%%%%%%%%%%%%%%%%%%%%%%%%%%%%%%%%%%%%

%%%%%%%%%%%%%%%%%%%%%%%%%%%%%%%%%%%%%%%%%%%%%%%%%%%%%%%%%%%%%%%%%%%%%%%%%%%
\begin{figure}
%\newlength{\picwidtha}
\setlength{\picwidtha}{3.0in}
 \begin{flushleft}
{\resizebox{\picwidtha}{!}{\includegraphics{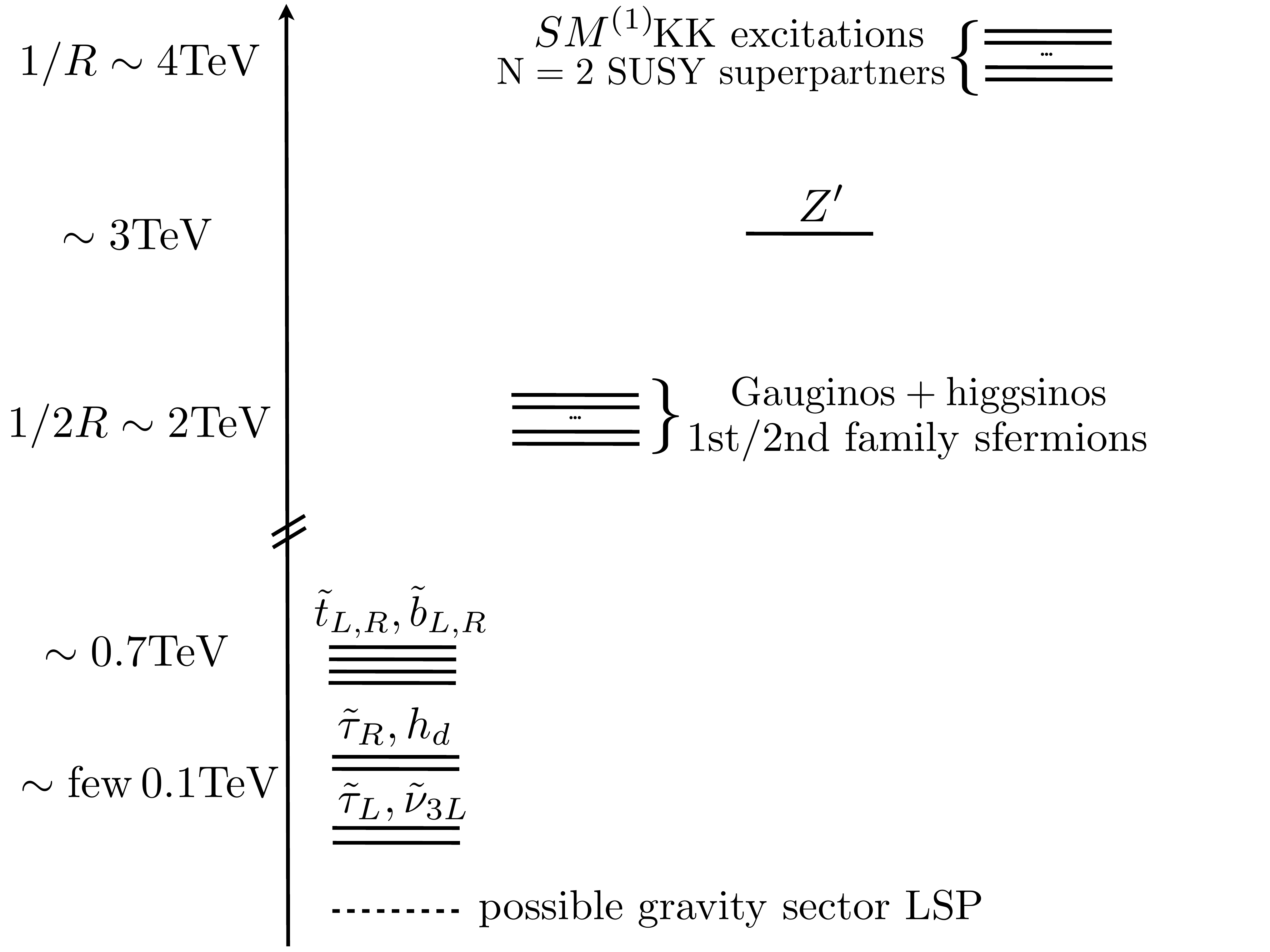}}}
 \end{flushleft}
\caption{\label{plot2}Schematic spectrum of new states of primary experimental interest.}
\end{figure}
%%%%%%%%%%%%%%%%%%%%%%%%%%%%%%%%%%%%%%%%%%%%%%%%%%%%%%%%%%%%%%%%%%%%%%%%%%%

The theory has a rich phenomenology, and a variety of new physics signatures are accessible to LHC14 in the low-fine-tuning parameter region.  Here we provide
just a brief summary of the main features~\cite{longpaper}.  The spectrum of new (non-gravitational) states is illustrated in Fig.~\ref{plot2}, where we have shown values with minimal
fine-tuning consistent with current bounds. 

The theory is mostly protected from precision, flavor and CP observables, although signatures are possible. While SUSY flavor problems are suppressed by the automatic near-degeneracy of 1st/2nd generation squarks and the near-Dirac masses of higgsinos and gauginos, KK gauge boson exchange can lead to deviations in kaon and especially $B_q$ mixing and rare decays depending on model-dependent details \cite{flavorpaper}.  The high scale of the KK states and $U(1)'$ sectors, $1/R\sim m_{Z'}\gtrsim 4 ~ \tev$ protects from EWPT\cite{Marandella:2004xm}. Higgs properties are automatically SM-like since only $H_u$ obtains a VEV, and the inert $H_d$ is easily made consistent with limits.

The presence of additional large gravitational dimensions constrains models of inflation and reheating. A detailed treatment is left to future work \cite{longpaper}, but we note that a small inflationary energy scale $V_I < M^4_5 \ll M_{\rm pl}^4$ can be consistent with recent evidence for tensor perturbations \cite{Ade:2014xna} if the extra gravitational dimensions and thus the corresponding 4D Planck mass are small during inflation, as in models of rapid asymmetric inflation \cite{ArkaniHamed:1999gq}.

%The extra large gravitational dimensions put interesting limits on reheating and inflationary models. A detailed treatment is left to future work \cite{longpaper}, but we note that recent evidence for tensor perturbations \cite{Ade:2014xna} can be consistent with models of rapid asymmetric inflation \cite{ArkaniHamed:1999gq}.

The leading signature of this model is sparticle production at the LHC and future colliders. Two important differences from generic natural SUSY phenomenology occur. First, $m_{\tilde g} \sim (3 \div 5) m_{\tilde t}$ arises without extra tuning, and tuning limits will likely be driven by direct production of 3rd generation sparticles, not gluino production. Second, the absence of a light higgsino leads to unusual stop and sbottom decay chains. The brane-localized 3rd generation slepton masses are dominantly from higher dimensional operators Eq.~(\ref{higgsHDO}), so either $\tilde{\tau}_R$ or $\tilde{\nu}_{\tau_L}$ could be the lightest ordinary superpartner (LOSP). Three-body decays of $\tilde{t}$ and $\tilde{b}$ to the LOSP can dilute missing energy signatures and lead to $\tau$-rich final states. Depending on the embedding of the 5D theory in the gravitational dimensions, the LOSP can be collider stable, or decay through prompt or displaced vertices to {\it extra-dimensional}-gravitini or other $R_p$-odd states in the bulk. In another variation, if $F_X$ is generated independently of SSSB, the associated goldstino remains light~\cite{Cheung:2011jq} and ordinary superpartners will decay directly to this state, mimicking more standard natural susy signatures.  For this short work we take the LHC8 bounds on $\tilde{t}\to t +{\rm MET}$ of $m_{\tilde t} \gtrsim 650 \gev$~\cite{Aad:2013ija,Chatrchyan:2013xna} as a guideline, but this can potentially be eased.

The mass and couplings of the new $Z'$ are restricted by the requirement $m_h \approx 126\gev$, suggesting this state is also likely to be accessible; 8~TeV limits require $m_{Z'} \gtrsim 3\tev$ \cite{ATLAS-CONF-2013-017,Chatrchyan:2012oaa}.

%%%%%%%%%%%%%%%%%%%%%%%%%%%%%%%%%%%%%%%%%%%%%%%%%%%%%%%%%%%%%%%%%%%%%%%%%%%
\begin{figure}
%\newlength{\picwidtha}
\setlength{\picwidtha}{3.5in}
 \begin{flushleft}
{\resizebox{\picwidtha}{!}{\includegraphics{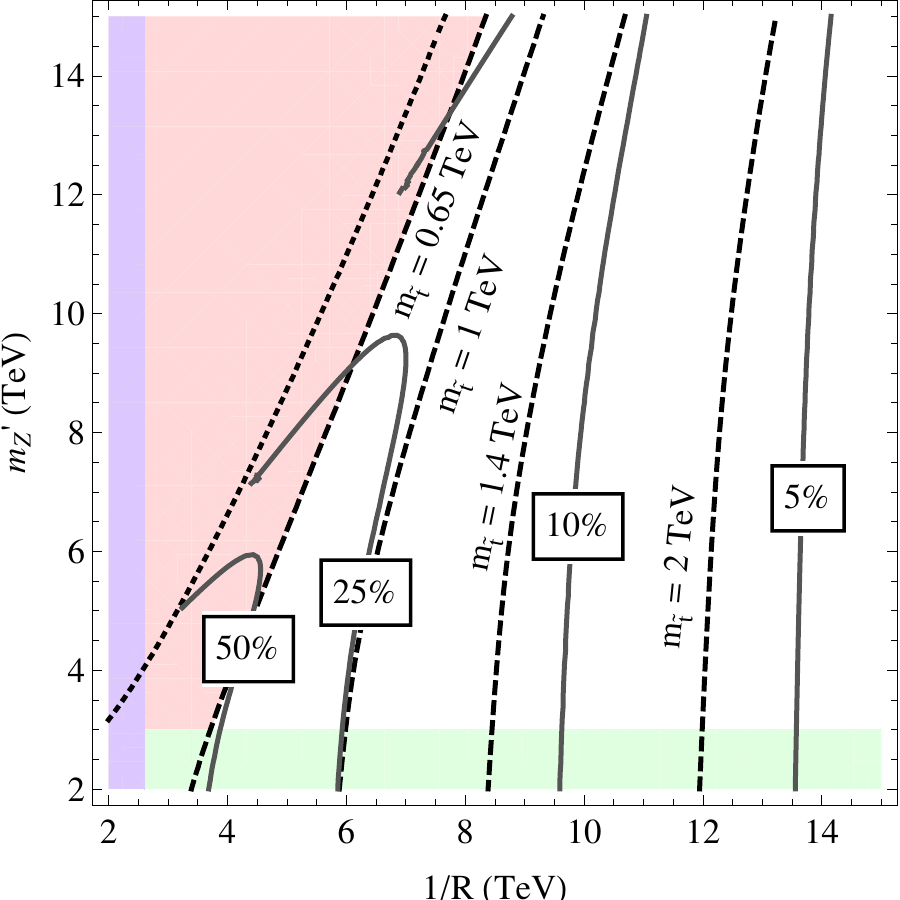}}}
 \end{flushleft}
\caption{\label{finetuning} Fine-tuning $\Delta^{-1}$ (solid lines) as function of $1/R$ and the $Z'$ mass,  Eq.(\ref{tuningeq}).   Iso-contours of stop mass are dashed. Limits from LHC8 searches for $\tilde{t}\to t+{\rm MET}$\cite{Aad:2013ija,Chatrchyan:2013xna} (red) and $Z'$ resonance searches \cite{ATLAS-CONF-2013-017,Chatrchyan:2012oaa} (green) are shaded. Subdominant limits $m_{\tilde{g}}\approx 1/(2R)\gtrsim 1.3 \tev$ from $\tilde{g}\to t \overline{t}/b\overline{b}+{\rm MET}$ searches (blue) are also shaded \cite{ATLAS-CONF-2013-061,Chatrchyan:2013iqa}. 
}
\end{figure}
%%%%%%%%%%%%%%%%%%%%%%%%%%%%%%%%%%%%%%%%%%%%%%%%%%%%%%%%%%%%%%%%%%%%%%%%%%%
	
The tuning of EWSB in this theory can be quantified by the sensitivity of $v$ to shifts at the scale $1/R$ of the stop mass (through the operator Eq.(\ref{higgsHDO})) and the $Z'$ mass,
\begin{gather}
	\Delta = \sqrt{\left(\frac{\partial \ln v^2}{\partial \ln {m^2_{\tilde t}}}\right)^2 +\left(\frac{\partial \ln v^2}{\partial \ln {m^2_{\tilde Z'}}}\right)^2},
\label{tuningeq}
\end{gather}
where for simplicity we set $m^2_{\tilde q_3}=m^2_{\tilde u_3}\equiv m^2_{\tilde t}$. The fine-tuning is shown in Fig.~\ref{finetuning}, where the stop mass has been fixed as a function of $1/R$ and $m_Z'$ to give successful EWSB. For $m_{Z'}\lesssim1.5/R$, the stop contribution is the dominant source of tuning. Remarkably at current LHC8 limits the theory is natural with a tuning of $\sim50\%$. LHC14 can discover stops at $m_{\tilde t}\sim1.2\gev$ \cite{ATL-PHYS-PUB-2013-011}, for which the theory is $\sim20\%$ tuned.
For $m_{\tilde{t}_1}\gtrsim 3.5\tev$, the tuning is still only at the few percent level and $m_h = 126~\gev$ might be obtained radiatively \cite{Feng:2013tvd} without the complications of an extra $U(1)'$ sector  -- an attractive target for a $100~\tev$ proton collider.

The production of KK excitations of SM particles would be an important signature of the extra-dimensional nature of this model, but their large mass $\sim 1/R$ and an approximate KK-parity make these particles difficult to reach at LHC14. Observing the near degeneracy of gauginos, higgsinos, and 1st/2nd generation sfermions would provide an alternative strong indication of the extra-dimensional nature of the theory.

In summary, we have presented a model where SSSB accompanied by a simple mechanism driving EWSB leads to a natural spectrum  consistent with Higgs properties and sparticle bounds with fine-tuning better than $\sim50\%$ even after LHC8 limits.  Variations involving different field content and localizations, including interplay with other mechanisms for driving EWSB in SSSB via different bc's \cite{Delgado:2001si,Delgado:2001xr} and quasi-localization of the stop \cite{Barbieri:2003kn,Barbieri:2002sw,Barbieri:2002uk,Barbieri:2000vh,Marti:2002ar} or Higgs \cite{Diego:2005mu,Diego:2006py,vonGersdorff:2007kk,Bhattacharyya:2012ct} deserve further attention as leading candidates for natural theories at LHC14 and future colliders. In an aesthetic direction, the extended gauge structure and extra dimensions suggest interesting possibilities for gauge unification in this model~\cite{longpaper}.

%%%%%%%%%%%%%%%%%%%%%%%%%%%%%%%%%%%%%%%%%%%%%%%%%%%%%%%%%%%%%%%%%%%%%%%%%%%%%
\begin{acknowledgments}
We thank N. Arkani-Hamed, A. Arvanitaki, M.~Baryakhtar, N.~Craig, I. Garcia~Garcia, T. Gherghetta, E.~Hardy, X.~Huang, D. E. Kaplan, K.~Van Tilburg, J.~Wacker, and Y.~Zhao for useful discussions.  We especially thank DEK for hospitality at JHU during a portion of this work.
This work was partially supported by ERC grant BSMOXFORD no. 228169. KH is supported by an NSF Graduate Research Fellowship under
Grant number DGE-0645962 and by the US DoE under contract DE-AC02-76SF00515. \end{acknowledgments}

%%%%%%%%%%%%%%%%%%%%%%%%%%%%%%%%%%%%%%%%%%%%%%%%%%%%%%%%%%%%%%%%%%%%%%%%%%%%%%
\bibliography{ScherkSchwarz} % Produces the bibliography via BibTeX.

\begin{thebibliography}{76}
\expandafter\ifx\csname natexlab\endcsname\relax\def\natexlab#1{#1}\fi
\expandafter\ifx\csname bibnamefont\endcsname\relax
  \def\bibnamefont#1{#1}\fi
\expandafter\ifx\csname bibfnamefont\endcsname\relax
  \def\bibfnamefont#1{#1}\fi
\expandafter\ifx\csname citenamefont\endcsname\relax
  \def\citenamefont#1{#1}\fi
\expandafter\ifx\csname url\endcsname\relax
  \def\url#1{\texttt{#1}}\fi
\expandafter\ifx\csname urlprefix\endcsname\relax\def\urlprefix{URL }\fi
\providecommand{\bibinfo}[2]{#2}
\providecommand{\eprint}[2][]{\url{#2}}

\bibitem[{\citenamefont{Gherghetta et~al.}(2013)\citenamefont{Gherghetta, von
  Harling, Medina, and Schmidt}}]{Gherghetta:2012gb}
\bibinfo{author}{\bibfnamefont{T.}~\bibnamefont{Gherghetta}},
  \bibinfo{author}{\bibfnamefont{B.}~\bibnamefont{von Harling}},
  \bibinfo{author}{\bibfnamefont{A.~D.} \bibnamefont{Medina}},
  \bibnamefont{and} \bibinfo{author}{\bibfnamefont{M.~A.}
  \bibnamefont{Schmidt}}, \bibinfo{journal}{JHEP}
  \textbf{\bibinfo{volume}{02}}, \bibinfo{pages}{032} (\bibinfo{year}{2013}),
  \eprint{1212.5243}.

\bibitem[{\citenamefont{Arvanitaki et~al.}(2014)\citenamefont{Arvanitaki,
  Baryakhtar, Huang, van Tilburg, and Villadoro}}]{Arvanitaki:2013yja}
\bibinfo{author}{\bibfnamefont{A.}~\bibnamefont{Arvanitaki}},
  \bibinfo{author}{\bibfnamefont{M.}~\bibnamefont{Baryakhtar}},
  \bibinfo{author}{\bibfnamefont{X.}~\bibnamefont{Huang}},
  \bibinfo{author}{\bibfnamefont{K.}~\bibnamefont{van Tilburg}},
  \bibnamefont{and}
  \bibinfo{author}{\bibfnamefont{G.}~\bibnamefont{Villadoro}},
  \bibinfo{journal}{JHEP} \textbf{\bibinfo{volume}{1403}}, \bibinfo{pages}{022}
  (\bibinfo{year}{2014}), \eprint{1309.3568}.

\bibitem[{\citenamefont{Hardy}(2013)}]{Hardy:2013ywa}
\bibinfo{author}{\bibfnamefont{E.}~\bibnamefont{Hardy}},
  \bibinfo{journal}{JHEP} \textbf{\bibinfo{volume}{1310}}, \bibinfo{pages}{133}
  (\bibinfo{year}{2013}), \eprint{1306.1534}.

\bibitem[{\citenamefont{Feng}(2013)}]{Feng:2013pwa}
\bibinfo{author}{\bibfnamefont{J.~L.} \bibnamefont{Feng}},
  \bibinfo{journal}{Ann.Rev.Nucl.Part.Sci.} \textbf{\bibinfo{volume}{63}},
  \bibinfo{pages}{351} (\bibinfo{year}{2013}), \eprint{1302.6587}.

\bibitem[{\citenamefont{Gherghetta et~al.}(2014)\citenamefont{Gherghetta, von
  Harling, Medina, and Schmidt}}]{Gherghetta:2014xea}
\bibinfo{author}{\bibfnamefont{T.}~\bibnamefont{Gherghetta}},
  \bibinfo{author}{\bibfnamefont{B.}~\bibnamefont{von Harling}},
  \bibinfo{author}{\bibfnamefont{A.~D.} \bibnamefont{Medina}},
  \bibnamefont{and} \bibinfo{author}{\bibfnamefont{M.~A.}
  \bibnamefont{Schmidt}} (\bibinfo{year}{2014}), \eprint{1401.8291}.

\bibitem[{\citenamefont{Fan and Reece}(2014)}]{Fan:2014txa}
\bibinfo{author}{\bibfnamefont{J.}~\bibnamefont{Fan}} \bibnamefont{and}
  \bibinfo{author}{\bibfnamefont{M.}~\bibnamefont{Reece}}
  (\bibinfo{year}{2014}), \eprint{1401.7671}.

\bibitem[{\citenamefont{Antoniadis et~al.}(1999)\citenamefont{Antoniadis,
  Dimopoulos, Pomarol, and Quiros}}]{Antoniadis:1998sd}
\bibinfo{author}{\bibfnamefont{I.}~\bibnamefont{Antoniadis}},
  \bibinfo{author}{\bibfnamefont{S.}~\bibnamefont{Dimopoulos}},
  \bibinfo{author}{\bibfnamefont{A.}~\bibnamefont{Pomarol}}, \bibnamefont{and}
  \bibinfo{author}{\bibfnamefont{M.}~\bibnamefont{Quiros}},
  \bibinfo{journal}{Nucl.Phys.} \textbf{\bibinfo{volume}{B544}},
  \bibinfo{pages}{503} (\bibinfo{year}{1999}), \eprint{hep-ph/9810410}.

\bibitem[{\citenamefont{Delgado et~al.}(1999)\citenamefont{Delgado, Pomarol,
  and Quiros}}]{Delgado:1998qr}
\bibinfo{author}{\bibfnamefont{A.}~\bibnamefont{Delgado}},
  \bibinfo{author}{\bibfnamefont{A.}~\bibnamefont{Pomarol}}, \bibnamefont{and}
  \bibinfo{author}{\bibfnamefont{M.}~\bibnamefont{Quiros}},
  \bibinfo{journal}{Phys.Rev.} \textbf{\bibinfo{volume}{D60}},
  \bibinfo{pages}{095008} (\bibinfo{year}{1999}), \eprint{hep-ph/9812489}.

\bibitem[{\citenamefont{Pomarol and Quiros}(1998)}]{Pomarol:1998sd}
\bibinfo{author}{\bibfnamefont{A.}~\bibnamefont{Pomarol}} \bibnamefont{and}
  \bibinfo{author}{\bibfnamefont{M.}~\bibnamefont{Quiros}},
  \bibinfo{journal}{Phys.Lett.} \textbf{\bibinfo{volume}{B438}},
  \bibinfo{pages}{255} (\bibinfo{year}{1998}), \eprint{hep-ph/9806263}.

\bibitem[{\citenamefont{Delgado and Quiros}(2001)}]{Delgado:2001si}
\bibinfo{author}{\bibfnamefont{A.}~\bibnamefont{Delgado}} \bibnamefont{and}
  \bibinfo{author}{\bibfnamefont{M.}~\bibnamefont{Quiros}},
  \bibinfo{journal}{Nucl.Phys.} \textbf{\bibinfo{volume}{B607}},
  \bibinfo{pages}{99} (\bibinfo{year}{2001}), \eprint{hep-ph/0103058}.

\bibitem[{\citenamefont{Delgado
  et~al.}(2001{\natexlab{a}})\citenamefont{Delgado, von Gersdorff, and
  Quiros}}]{Delgado:2001xr}
\bibinfo{author}{\bibfnamefont{A.}~\bibnamefont{Delgado}},
  \bibinfo{author}{\bibfnamefont{G.}~\bibnamefont{von Gersdorff}},
  \bibnamefont{and} \bibinfo{author}{\bibfnamefont{M.}~\bibnamefont{Quiros}},
  \bibinfo{journal}{Nucl.Phys.} \textbf{\bibinfo{volume}{B613}},
  \bibinfo{pages}{49} (\bibinfo{year}{2001}{\natexlab{a}}),
  \eprint{hep-ph/0107233}.

\bibitem[{\citenamefont{Barbieri
  et~al.}(2003{\natexlab{a}})\citenamefont{Barbieri, Marandella, and
  Papucci}}]{Barbieri:2003kn}
\bibinfo{author}{\bibfnamefont{R.}~\bibnamefont{Barbieri}},
  \bibinfo{author}{\bibfnamefont{G.}~\bibnamefont{Marandella}},
  \bibnamefont{and} \bibinfo{author}{\bibfnamefont{M.}~\bibnamefont{Papucci}},
  \bibinfo{journal}{Nucl.Phys.} \textbf{\bibinfo{volume}{B668}},
  \bibinfo{pages}{273} (\bibinfo{year}{2003}{\natexlab{a}}),
  \eprint{hep-ph/0305044}.

\bibitem[{\citenamefont{Barbieri
  et~al.}(2003{\natexlab{b}})\citenamefont{Barbieri, Hall, Marandella, Nomura,
  Okui et~al.}}]{Barbieri:2002sw}
\bibinfo{author}{\bibfnamefont{R.}~\bibnamefont{Barbieri}},
  \bibinfo{author}{\bibfnamefont{L.~J.} \bibnamefont{Hall}},
  \bibinfo{author}{\bibfnamefont{G.}~\bibnamefont{Marandella}},
  \bibinfo{author}{\bibfnamefont{Y.}~\bibnamefont{Nomura}},
  \bibinfo{author}{\bibfnamefont{T.}~\bibnamefont{Okui}}, \bibnamefont{et~al.},
  \bibinfo{journal}{Nucl.Phys.} \textbf{\bibinfo{volume}{B663}},
  \bibinfo{pages}{141} (\bibinfo{year}{2003}{\natexlab{b}}),
  \eprint{hep-ph/0208153}.

\bibitem[{\citenamefont{Barbieri
  et~al.}(2002{\natexlab{a}})\citenamefont{Barbieri, Marandella, and
  Papucci}}]{Barbieri:2002uk}
\bibinfo{author}{\bibfnamefont{R.}~\bibnamefont{Barbieri}},
  \bibinfo{author}{\bibfnamefont{G.}~\bibnamefont{Marandella}},
  \bibnamefont{and} \bibinfo{author}{\bibfnamefont{M.}~\bibnamefont{Papucci}},
  \bibinfo{journal}{Phys.Rev.} \textbf{\bibinfo{volume}{D66}},
  \bibinfo{pages}{095003} (\bibinfo{year}{2002}{\natexlab{a}}),
  \eprint{hep-ph/0205280}.

\bibitem[{\citenamefont{Barbieri et~al.}(2001)\citenamefont{Barbieri, Hall, and
  Nomura}}]{Barbieri:2000vh}
\bibinfo{author}{\bibfnamefont{R.}~\bibnamefont{Barbieri}},
  \bibinfo{author}{\bibfnamefont{L.~J.} \bibnamefont{Hall}}, \bibnamefont{and}
  \bibinfo{author}{\bibfnamefont{Y.}~\bibnamefont{Nomura}},
  \bibinfo{journal}{Phys.Rev.} \textbf{\bibinfo{volume}{D63}},
  \bibinfo{pages}{105007} (\bibinfo{year}{2001}), \eprint{hep-ph/0011311}.

\bibitem[{\citenamefont{Marti and Pomarol}(2002)}]{Marti:2002ar}
\bibinfo{author}{\bibfnamefont{D.}~\bibnamefont{Marti}} \bibnamefont{and}
  \bibinfo{author}{\bibfnamefont{A.}~\bibnamefont{Pomarol}},
  \bibinfo{journal}{Phys.Rev.} \textbf{\bibinfo{volume}{D66}},
  \bibinfo{pages}{125005} (\bibinfo{year}{2002}), \eprint{hep-ph/0205034}.

\bibitem[{\citenamefont{Diego et~al.}(2005)\citenamefont{Diego, von Gersdorff,
  and Quiros}}]{Diego:2005mu}
\bibinfo{author}{\bibfnamefont{D.}~\bibnamefont{Diego}},
  \bibinfo{author}{\bibfnamefont{G.}~\bibnamefont{von Gersdorff}},
  \bibnamefont{and} \bibinfo{author}{\bibfnamefont{M.}~\bibnamefont{Quiros}},
  \bibinfo{journal}{JHEP} \textbf{\bibinfo{volume}{0511}}, \bibinfo{pages}{008}
  (\bibinfo{year}{2005}), \eprint{hep-ph/0505244}.

\bibitem[{\citenamefont{Diego et~al.}(2006)\citenamefont{Diego, von Gersdorff,
  and Quiros}}]{Diego:2006py}
\bibinfo{author}{\bibfnamefont{D.}~\bibnamefont{Diego}},
  \bibinfo{author}{\bibfnamefont{G.}~\bibnamefont{von Gersdorff}},
  \bibnamefont{and} \bibinfo{author}{\bibfnamefont{M.}~\bibnamefont{Quiros}},
  \bibinfo{journal}{Phys.Rev.} \textbf{\bibinfo{volume}{D74}},
  \bibinfo{pages}{055004} (\bibinfo{year}{2006}), \eprint{hep-ph/0605024}.

\bibitem[{\citenamefont{von Gersdorff}(2007)}]{vonGersdorff:2007kk}
\bibinfo{author}{\bibfnamefont{G.}~\bibnamefont{von Gersdorff}},
  \bibinfo{journal}{Mod.Phys.Lett.} \textbf{\bibinfo{volume}{A22}},
  \bibinfo{pages}{385} (\bibinfo{year}{2007}), \eprint{hep-ph/0701256}.

\bibitem[{\citenamefont{Bhattacharyya and Ray}(2012)}]{Bhattacharyya:2012ct}
\bibinfo{author}{\bibfnamefont{G.}~\bibnamefont{Bhattacharyya}}
  \bibnamefont{and} \bibinfo{author}{\bibfnamefont{T.~S.} \bibnamefont{Ray}},
  \bibinfo{journal}{JHEP} \textbf{\bibinfo{volume}{1205}}, \bibinfo{pages}{022}
  (\bibinfo{year}{2012}), \eprint{1201.1131}.

\bibitem[{\citenamefont{Dimopoulos and Giudice}(1995)}]{Dimopoulos:1995mi}
\bibinfo{author}{\bibfnamefont{S.}~\bibnamefont{Dimopoulos}} \bibnamefont{and}
  \bibinfo{author}{\bibfnamefont{G.}~\bibnamefont{Giudice}},
  \bibinfo{journal}{Phys.Lett.} \textbf{\bibinfo{volume}{B357}},
  \bibinfo{pages}{573} (\bibinfo{year}{1995}), \eprint{hep-ph/9507282}.

\bibitem[{\citenamefont{Pomarol and Tommasini}(1996)}]{Pomarol:1995xc}
\bibinfo{author}{\bibfnamefont{A.}~\bibnamefont{Pomarol}} \bibnamefont{and}
  \bibinfo{author}{\bibfnamefont{D.}~\bibnamefont{Tommasini}},
  \bibinfo{journal}{Nucl.Phys.} \textbf{\bibinfo{volume}{B466}},
  \bibinfo{pages}{3} (\bibinfo{year}{1996}), \eprint{hep-ph/9507462}.

\bibitem[{\citenamefont{Cohen et~al.}(1996)\citenamefont{Cohen, Kaplan, and
  Nelson}}]{Cohen:1996vb}
\bibinfo{author}{\bibfnamefont{A.~G.} \bibnamefont{Cohen}},
  \bibinfo{author}{\bibfnamefont{D.}~\bibnamefont{Kaplan}}, \bibnamefont{and}
  \bibinfo{author}{\bibfnamefont{A.}~\bibnamefont{Nelson}},
  \bibinfo{journal}{Phys.Lett.} \textbf{\bibinfo{volume}{B388}},
  \bibinfo{pages}{588} (\bibinfo{year}{1996}), \eprint{hep-ph/9607394}.

\bibitem[{\citenamefont{Antoniadis
  et~al.}(1998{\natexlab{a}})\citenamefont{Antoniadis, Dimopoulos, and
  Dvali}}]{Antoniadis:1997zg}
\bibinfo{author}{\bibfnamefont{I.}~\bibnamefont{Antoniadis}},
  \bibinfo{author}{\bibfnamefont{S.}~\bibnamefont{Dimopoulos}},
  \bibnamefont{and} \bibinfo{author}{\bibfnamefont{G.}~\bibnamefont{Dvali}},
  \bibinfo{journal}{Nucl.Phys.} \textbf{\bibinfo{volume}{B516}},
  \bibinfo{pages}{70} (\bibinfo{year}{1998}{\natexlab{a}}),
  \eprint{hep-ph/9710204}.

\bibitem[{\citenamefont{Barbieri
  et~al.}(2002{\natexlab{b}})\citenamefont{Barbieri, Hall, and
  Nomura}}]{Barbieri:2001dm}
\bibinfo{author}{\bibfnamefont{R.}~\bibnamefont{Barbieri}},
  \bibinfo{author}{\bibfnamefont{L.~J.} \bibnamefont{Hall}}, \bibnamefont{and}
  \bibinfo{author}{\bibfnamefont{Y.}~\bibnamefont{Nomura}},
  \bibinfo{journal}{Nucl.Phys.} \textbf{\bibinfo{volume}{B624}},
  \bibinfo{pages}{63} (\bibinfo{year}{2002}{\natexlab{b}}),
  \eprint{hep-th/0107004}.

\bibitem[{\citenamefont{Delgado
  et~al.}(2001{\natexlab{b}})\citenamefont{Delgado, von Gersdorff, John, and
  Quiros}}]{Delgado:2001ex}
\bibinfo{author}{\bibfnamefont{A.}~\bibnamefont{Delgado}},
  \bibinfo{author}{\bibfnamefont{G.}~\bibnamefont{von Gersdorff}},
  \bibinfo{author}{\bibfnamefont{P.}~\bibnamefont{John}}, \bibnamefont{and}
  \bibinfo{author}{\bibfnamefont{M.}~\bibnamefont{Quiros}},
  \bibinfo{journal}{Phys.Lett.} \textbf{\bibinfo{volume}{B517}},
  \bibinfo{pages}{445} (\bibinfo{year}{2001}{\natexlab{b}}),
  \eprint{hep-ph/0104112}.

\bibitem[{\citenamefont{Contino and Pilo}(2001)}]{Contino:2001gz}
\bibinfo{author}{\bibfnamefont{R.}~\bibnamefont{Contino}} \bibnamefont{and}
  \bibinfo{author}{\bibfnamefont{L.}~\bibnamefont{Pilo}},
  \bibinfo{journal}{Phys.Lett.} \textbf{\bibinfo{volume}{B523}},
  \bibinfo{pages}{347} (\bibinfo{year}{2001}), \eprint{hep-ph/0104130}.

\bibitem[{\citenamefont{Weiner}(2001)}]{Weiner:2001ui}
\bibinfo{author}{\bibfnamefont{N.}~\bibnamefont{Weiner}}
  (\bibinfo{year}{2001}), \eprint{hep-ph/0106021}.

\bibitem[{\citenamefont{Kim}(2002)}]{Kim:2001re}
\bibinfo{author}{\bibfnamefont{H.~D.} \bibnamefont{Kim}},
  \bibinfo{journal}{Phys.Rev.} \textbf{\bibinfo{volume}{D65}},
  \bibinfo{pages}{105021} (\bibinfo{year}{2002}), \eprint{hep-th/0109101}.

\bibitem[{\citenamefont{Puchwein and Kunszt}(2004)}]{Puchwein:2003jq}
\bibinfo{author}{\bibfnamefont{M.}~\bibnamefont{Puchwein}} \bibnamefont{and}
  \bibinfo{author}{\bibfnamefont{Z.}~\bibnamefont{Kunszt}},
  \bibinfo{journal}{Annals Phys.} \textbf{\bibinfo{volume}{311}},
  \bibinfo{pages}{288} (\bibinfo{year}{2004}), \eprint{hep-th/0309069}.

\bibitem[{\citenamefont{Davies et~al.}(2011)\citenamefont{Davies,
  March-Russell, and McCullough}}]{Davies:2011mp}
\bibinfo{author}{\bibfnamefont{R.}~\bibnamefont{Davies}},
  \bibinfo{author}{\bibfnamefont{J.}~\bibnamefont{March-Russell}},
  \bibnamefont{and}
  \bibinfo{author}{\bibfnamefont{M.}~\bibnamefont{McCullough}},
  \bibinfo{journal}{JHEP} \textbf{\bibinfo{volume}{1104}}, \bibinfo{pages}{108}
  (\bibinfo{year}{2011}), \eprint{1103.1647}.

\bibitem[{\citenamefont{Scherk and
  Schwarz}(1979{\natexlab{a}})}]{Scherk:1978ta}
\bibinfo{author}{\bibfnamefont{J.}~\bibnamefont{Scherk}} \bibnamefont{and}
  \bibinfo{author}{\bibfnamefont{J.~H.} \bibnamefont{Schwarz}},
  \bibinfo{journal}{Phys.Lett.} \textbf{\bibinfo{volume}{B82}},
  \bibinfo{pages}{60} (\bibinfo{year}{1979}{\natexlab{a}}).

\bibitem[{\citenamefont{Scherk and
  Schwarz}(1979{\natexlab{b}})}]{Scherk:1979zr}
\bibinfo{author}{\bibfnamefont{J.}~\bibnamefont{Scherk}} \bibnamefont{and}
  \bibinfo{author}{\bibfnamefont{J.~H.} \bibnamefont{Schwarz}},
  \bibinfo{journal}{Nucl.Phys.} \textbf{\bibinfo{volume}{B153}},
  \bibinfo{pages}{61} (\bibinfo{year}{1979}{\natexlab{b}}).

\bibitem[{\citenamefont{Marcus et~al.}(1983)\citenamefont{Marcus, Sagnotti, and
  Siegel}}]{Marcus:1983wb}
\bibinfo{author}{\bibfnamefont{N.}~\bibnamefont{Marcus}},
  \bibinfo{author}{\bibfnamefont{A.}~\bibnamefont{Sagnotti}}, \bibnamefont{and}
  \bibinfo{author}{\bibfnamefont{W.}~\bibnamefont{Siegel}},
  \bibinfo{journal}{Nucl.Phys.} \textbf{\bibinfo{volume}{B224}},
  \bibinfo{pages}{159} (\bibinfo{year}{1983}).

\bibitem[{\citenamefont{Arkani-Hamed
  et~al.}(2002{\natexlab{a}})\citenamefont{Arkani-Hamed, Gregoire, and
  Wacker}}]{ArkaniHamed:2001tb}
\bibinfo{author}{\bibfnamefont{N.}~\bibnamefont{Arkani-Hamed}},
  \bibinfo{author}{\bibfnamefont{T.}~\bibnamefont{Gregoire}}, \bibnamefont{and}
  \bibinfo{author}{\bibfnamefont{J.~G.} \bibnamefont{Wacker}},
  \bibinfo{journal}{JHEP} \textbf{\bibinfo{volume}{0203}}, \bibinfo{pages}{055}
  (\bibinfo{year}{2002}{\natexlab{a}}), \eprint{hep-th/0101233}.

\bibitem[{\citenamefont{Marti and Pomarol}(2001)}]{Marti:2001iw}
\bibinfo{author}{\bibfnamefont{D.}~\bibnamefont{Marti}} \bibnamefont{and}
  \bibinfo{author}{\bibfnamefont{A.}~\bibnamefont{Pomarol}},
  \bibinfo{journal}{Phys.Rev.} \textbf{\bibinfo{volume}{D64}},
  \bibinfo{pages}{105025} (\bibinfo{year}{2001}), \eprint{hep-th/0106256}.

\bibitem[{\citenamefont{Hebecker}(2002)}]{Hebecker:2001ke}
\bibinfo{author}{\bibfnamefont{A.}~\bibnamefont{Hebecker}},
  \bibinfo{journal}{Nucl.Phys.} \textbf{\bibinfo{volume}{B632}},
  \bibinfo{pages}{101} (\bibinfo{year}{2002}), \eprint{hep-ph/0112230}.

\bibitem[{\citenamefont{Linch et~al.}(2003)\citenamefont{Linch, Luty, and
  Phillips}}]{Linch:2002wg}
\bibinfo{author}{\bibfnamefont{I.}~\bibnamefont{Linch},
  \bibfnamefont{William~Divine}}, \bibinfo{author}{\bibfnamefont{M.~A.}
  \bibnamefont{Luty}}, \bibnamefont{and}
  \bibinfo{author}{\bibfnamefont{J.}~\bibnamefont{Phillips}},
  \bibinfo{journal}{Phys.Rev.} \textbf{\bibinfo{volume}{D68}},
  \bibinfo{pages}{025008} (\bibinfo{year}{2003}), \eprint{hep-th/0209060}.

\bibitem[{\citenamefont{Dimopoulos et~al.}(In
  preparation{\natexlab{a}})\citenamefont{Dimopoulos, Howe, and
  March-Russell}}]{longpaper}
\bibinfo{author}{\bibfnamefont{S.}~\bibnamefont{Dimopoulos}},
  \bibinfo{author}{\bibfnamefont{K.}~\bibnamefont{Howe}}, \bibnamefont{and}
  \bibinfo{author}{\bibfnamefont{J.}~\bibnamefont{March-Russell}}
  (\bibinfo{year}{In preparation}{\natexlab{a}}).

\bibitem[{\citenamefont{Muck et~al.}(2005)\citenamefont{Muck, Nilse, Pilaftsis,
  and Ruckl}}]{Muck:2004br}
\bibinfo{author}{\bibfnamefont{A.}~\bibnamefont{Muck}},
  \bibinfo{author}{\bibfnamefont{L.}~\bibnamefont{Nilse}},
  \bibinfo{author}{\bibfnamefont{A.}~\bibnamefont{Pilaftsis}},
  \bibnamefont{and} \bibinfo{author}{\bibfnamefont{R.}~\bibnamefont{Ruckl}},
  \bibinfo{journal}{Phys.Rev.} \textbf{\bibinfo{volume}{D71}},
  \bibinfo{pages}{066004} (\bibinfo{year}{2005}), \eprint{hep-ph/0411258}.

\bibitem[{\citenamefont{Chivukula et~al.}(2003)\citenamefont{Chivukula, Dicus,
  He, and Nandi}}]{Chivukula:2003kq}
\bibinfo{author}{\bibfnamefont{R.~S.} \bibnamefont{Chivukula}},
  \bibinfo{author}{\bibfnamefont{D.~A.} \bibnamefont{Dicus}},
  \bibinfo{author}{\bibfnamefont{H.-J.} \bibnamefont{He}}, \bibnamefont{and}
  \bibinfo{author}{\bibfnamefont{S.}~\bibnamefont{Nandi}},
  \bibinfo{journal}{Phys.Lett.} \textbf{\bibinfo{volume}{B562}},
  \bibinfo{pages}{109} (\bibinfo{year}{2003}), \eprint{hep-ph/0302263}.

\bibitem[{\citenamefont{Marandella and Papucci}(2005)}]{Marandella:2004xm}
\bibinfo{author}{\bibfnamefont{G.}~\bibnamefont{Marandella}} \bibnamefont{and}
  \bibinfo{author}{\bibfnamefont{M.}~\bibnamefont{Papucci}},
  \bibinfo{journal}{Phys.Rev.} \textbf{\bibinfo{volume}{D71}},
  \bibinfo{pages}{055010} (\bibinfo{year}{2005}), \eprint{hep-ph/0407030}.

\bibitem[{\citenamefont{Arkani-Hamed et~al.}(1999)\citenamefont{Arkani-Hamed,
  Dimopoulos, and Dvali}}]{ArkaniHamed:1998nn}
\bibinfo{author}{\bibfnamefont{N.}~\bibnamefont{Arkani-Hamed}},
  \bibinfo{author}{\bibfnamefont{S.}~\bibnamefont{Dimopoulos}},
  \bibnamefont{and} \bibinfo{author}{\bibfnamefont{G.}~\bibnamefont{Dvali}},
  \bibinfo{journal}{Phys.Rev.} \textbf{\bibinfo{volume}{D59}},
  \bibinfo{pages}{086004} (\bibinfo{year}{1999}), \eprint{hep-ph/9807344}.

\bibitem[{\citenamefont{Antoniadis
  et~al.}(1998{\natexlab{b}})\citenamefont{Antoniadis, Arkani-Hamed,
  Dimopoulos, and Dvali}}]{Antoniadis:1998ig}
\bibinfo{author}{\bibfnamefont{I.}~\bibnamefont{Antoniadis}},
  \bibinfo{author}{\bibfnamefont{N.}~\bibnamefont{Arkani-Hamed}},
  \bibinfo{author}{\bibfnamefont{S.}~\bibnamefont{Dimopoulos}},
  \bibnamefont{and} \bibinfo{author}{\bibfnamefont{G.}~\bibnamefont{Dvali}},
  \bibinfo{journal}{Phys.Lett.} \textbf{\bibinfo{volume}{B436}},
  \bibinfo{pages}{257} (\bibinfo{year}{1998}{\natexlab{b}}),
  \eprint{hep-ph/9804398}.

\bibitem[{\citenamefont{Arkani-Hamed
  et~al.}(2002{\natexlab{b}})\citenamefont{Arkani-Hamed, Dimopoulos, Dvali, and
  March-Russell}}]{ArkaniHamed:1998vp}
\bibinfo{author}{\bibfnamefont{N.}~\bibnamefont{Arkani-Hamed}},
  \bibinfo{author}{\bibfnamefont{S.}~\bibnamefont{Dimopoulos}},
  \bibinfo{author}{\bibfnamefont{G.}~\bibnamefont{Dvali}}, \bibnamefont{and}
  \bibinfo{author}{\bibfnamefont{J.}~\bibnamefont{March-Russell}},
  \bibinfo{journal}{Phys.Rev.} \textbf{\bibinfo{volume}{D65}},
  \bibinfo{pages}{024032} (\bibinfo{year}{2002}{\natexlab{b}}),
  \eprint{hep-ph/9811448}.

\bibitem[{\citenamefont{Argyres et~al.}(1998)\citenamefont{Argyres, Dimopoulos,
  and March-Russell}}]{Argyres:1998qn}
\bibinfo{author}{\bibfnamefont{P.~C.} \bibnamefont{Argyres}},
  \bibinfo{author}{\bibfnamefont{S.}~\bibnamefont{Dimopoulos}},
  \bibnamefont{and}
  \bibinfo{author}{\bibfnamefont{J.}~\bibnamefont{March-Russell}},
  \bibinfo{journal}{Phys.Lett.} \textbf{\bibinfo{volume}{B441}},
  \bibinfo{pages}{96} (\bibinfo{year}{1998}), \eprint{hep-th/9808138}.

\bibitem[{\citenamefont{Antoniadis et~al.}(2012)\citenamefont{Antoniadis,
  Arvanitaki, Dimopoulos, and Giveon}}]{Antoniadis:2011qw}
\bibinfo{author}{\bibfnamefont{I.}~\bibnamefont{Antoniadis}},
  \bibinfo{author}{\bibfnamefont{A.}~\bibnamefont{Arvanitaki}},
  \bibinfo{author}{\bibfnamefont{S.}~\bibnamefont{Dimopoulos}},
  \bibnamefont{and} \bibinfo{author}{\bibfnamefont{A.}~\bibnamefont{Giveon}},
  \bibinfo{journal}{Phys.Rev.Lett.} \textbf{\bibinfo{volume}{108}},
  \bibinfo{pages}{081602} (\bibinfo{year}{2012}), \eprint{1102.4043}.

\bibitem[{\citenamefont{Luty and Okada}(2003)}]{Luty:2002hj}
\bibinfo{author}{\bibfnamefont{M.~A.} \bibnamefont{Luty}} \bibnamefont{and}
  \bibinfo{author}{\bibfnamefont{N.}~\bibnamefont{Okada}},
  \bibinfo{journal}{JHEP} \textbf{\bibinfo{volume}{0304}}, \bibinfo{pages}{050}
  (\bibinfo{year}{2003}), \eprint{hep-th/0209178}.

\bibitem[{\citenamefont{Kaplan and Weiner}(2001)}]{Kaplan:2001cg}
\bibinfo{author}{\bibfnamefont{D.~E.} \bibnamefont{Kaplan}} \bibnamefont{and}
  \bibinfo{author}{\bibfnamefont{N.}~\bibnamefont{Weiner}}
  (\bibinfo{year}{2001}), \eprint{hep-ph/0108001}.

\bibitem[{\citenamefont{Ponton and Poppitz}(2001)}]{Ponton:2001hq}
\bibinfo{author}{\bibfnamefont{E.}~\bibnamefont{Ponton}} \bibnamefont{and}
  \bibinfo{author}{\bibfnamefont{E.}~\bibnamefont{Poppitz}},
  \bibinfo{journal}{JHEP} \textbf{\bibinfo{volume}{0106}}, \bibinfo{pages}{019}
  (\bibinfo{year}{2001}), \eprint{hep-ph/0105021}.

\bibitem[{\citenamefont{von Gersdorff et~al.}(2004)\citenamefont{von Gersdorff,
  Quiros, and Riotto}}]{vonGersdorff:2003rq}
\bibinfo{author}{\bibfnamefont{G.}~\bibnamefont{von Gersdorff}},
  \bibinfo{author}{\bibfnamefont{M.}~\bibnamefont{Quiros}}, \bibnamefont{and}
  \bibinfo{author}{\bibfnamefont{A.}~\bibnamefont{Riotto}},
  \bibinfo{journal}{Nucl.Phys.} \textbf{\bibinfo{volume}{B689}},
  \bibinfo{pages}{76} (\bibinfo{year}{2004}), \eprint{hep-th/0310190}.

\bibitem[{\citenamefont{Rattazzi et~al.}(2003)\citenamefont{Rattazzi, Scrucca,
  and Strumia}}]{Rattazzi:2003rj}
\bibinfo{author}{\bibfnamefont{R.}~\bibnamefont{Rattazzi}},
  \bibinfo{author}{\bibfnamefont{C.~A.} \bibnamefont{Scrucca}},
  \bibnamefont{and} \bibinfo{author}{\bibfnamefont{A.}~\bibnamefont{Strumia}},
  \bibinfo{journal}{Nucl.Phys.} \textbf{\bibinfo{volume}{B674}},
  \bibinfo{pages}{171} (\bibinfo{year}{2003}), \eprint{hep-th/0305184}.

\bibitem[{\citenamefont{von Gersdorff and
  Hebecker}(2005)}]{vonGersdorff:2005ce}
\bibinfo{author}{\bibfnamefont{G.}~\bibnamefont{von Gersdorff}}
  \bibnamefont{and} \bibinfo{author}{\bibfnamefont{A.}~\bibnamefont{Hebecker}},
  \bibinfo{journal}{Nucl.Phys.} \textbf{\bibinfo{volume}{B720}},
  \bibinfo{pages}{211} (\bibinfo{year}{2005}), \eprint{hep-th/0504002}.

\bibitem[{\citenamefont{Dudas and Quiros}(2005)}]{Dudas:2005vna}
\bibinfo{author}{\bibfnamefont{E.}~\bibnamefont{Dudas}} \bibnamefont{and}
  \bibinfo{author}{\bibfnamefont{M.}~\bibnamefont{Quiros}},
  \bibinfo{journal}{Nucl.Phys.} \textbf{\bibinfo{volume}{B721}},
  \bibinfo{pages}{309} (\bibinfo{year}{2005}), \eprint{hep-th/0503157}.

\bibitem[{\citenamefont{Ghilencea et~al.}(2001)\citenamefont{Ghilencea,
  Groot~Nibbelink, and Nilles}}]{Ghilencea:2001bw}
\bibinfo{author}{\bibfnamefont{D.}~\bibnamefont{Ghilencea}},
  \bibinfo{author}{\bibfnamefont{S.}~\bibnamefont{Groot~Nibbelink}},
  \bibnamefont{and} \bibinfo{author}{\bibfnamefont{H.~P.}
  \bibnamefont{Nilles}}, \bibinfo{journal}{Nucl.Phys.}
  \textbf{\bibinfo{volume}{B619}}, \bibinfo{pages}{385} (\bibinfo{year}{2001}),
  \eprint{hep-th/0108184}.

\bibitem[{\citenamefont{Barbieri
  et~al.}(2002{\natexlab{c}})\citenamefont{Barbieri, Contino, Creminelli,
  Rattazzi, and Scrucca}}]{Barbieri:2002ic}
\bibinfo{author}{\bibfnamefont{R.}~\bibnamefont{Barbieri}},
  \bibinfo{author}{\bibfnamefont{R.}~\bibnamefont{Contino}},
  \bibinfo{author}{\bibfnamefont{P.}~\bibnamefont{Creminelli}},
  \bibinfo{author}{\bibfnamefont{R.}~\bibnamefont{Rattazzi}}, \bibnamefont{and}
  \bibinfo{author}{\bibfnamefont{C.}~\bibnamefont{Scrucca}},
  \bibinfo{journal}{Phys.Rev.} \textbf{\bibinfo{volume}{D66}},
  \bibinfo{pages}{024025} (\bibinfo{year}{2002}{\natexlab{c}}),
  \eprint{hep-th/0203039}.

\bibitem[{\citenamefont{Cirelli et~al.}(2006)\citenamefont{Cirelli, Fornengo,
  and Strumia}}]{Cirelli:2005uq}
\bibinfo{author}{\bibfnamefont{M.}~\bibnamefont{Cirelli}},
  \bibinfo{author}{\bibfnamefont{N.}~\bibnamefont{Fornengo}}, \bibnamefont{and}
  \bibinfo{author}{\bibfnamefont{A.}~\bibnamefont{Strumia}},
  \bibinfo{journal}{Nucl.Phys.} \textbf{\bibinfo{volume}{B753}},
  \bibinfo{pages}{178} (\bibinfo{year}{2006}), \eprint{hep-ph/0512090}.

\bibitem[{\citenamefont{Barbieri et~al.}(2006)\citenamefont{Barbieri, Hall, and
  Rychkov}}]{Barbieri:2006dq}
\bibinfo{author}{\bibfnamefont{R.}~\bibnamefont{Barbieri}},
  \bibinfo{author}{\bibfnamefont{L.~J.} \bibnamefont{Hall}}, \bibnamefont{and}
  \bibinfo{author}{\bibfnamefont{V.~S.} \bibnamefont{Rychkov}},
  \bibinfo{journal}{Phys.Rev.} \textbf{\bibinfo{volume}{D74}},
  \bibinfo{pages}{015007} (\bibinfo{year}{2006}), \eprint{hep-ph/0603188}.

\bibitem[{\citenamefont{Ma}(2006)}]{Ma:2006km}
\bibinfo{author}{\bibfnamefont{E.}~\bibnamefont{Ma}},
  \bibinfo{journal}{Phys.Rev.} \textbf{\bibinfo{volume}{D73}},
  \bibinfo{pages}{077301} (\bibinfo{year}{2006}), \eprint{hep-ph/0601225}.

\bibitem[{\citenamefont{Lopez~Honorez et~al.}(2007)\citenamefont{Lopez~Honorez,
  Nezri, Oliver, and Tytgat}}]{LopezHonorez:2006gr}
\bibinfo{author}{\bibfnamefont{L.}~\bibnamefont{Lopez~Honorez}},
  \bibinfo{author}{\bibfnamefont{E.}~\bibnamefont{Nezri}},
  \bibinfo{author}{\bibfnamefont{J.~F.} \bibnamefont{Oliver}},
  \bibnamefont{and} \bibinfo{author}{\bibfnamefont{M.~H.}
  \bibnamefont{Tytgat}}, \bibinfo{journal}{JCAP}
  \textbf{\bibinfo{volume}{0702}}, \bibinfo{pages}{028} (\bibinfo{year}{2007}),
  \eprint{hep-ph/0612275}.

\bibitem[{\citenamefont{Feng et~al.}(2013)\citenamefont{Feng, Kant, Profumo,
  and Sanford}}]{Feng:2013tvd}
\bibinfo{author}{\bibfnamefont{J.~L.} \bibnamefont{Feng}},
  \bibinfo{author}{\bibfnamefont{P.}~\bibnamefont{Kant}},
  \bibinfo{author}{\bibfnamefont{S.}~\bibnamefont{Profumo}}, \bibnamefont{and}
  \bibinfo{author}{\bibfnamefont{D.}~\bibnamefont{Sanford}},
  \bibinfo{journal}{Phys.Rev.Lett.} \textbf{\bibinfo{volume}{111}},
  \bibinfo{pages}{131802} (\bibinfo{year}{2013}), \eprint{1306.2318}.

\bibitem[{\citenamefont{Delgado and Quiros}(2000)}]{Delgado:2000fs}
\bibinfo{author}{\bibfnamefont{A.}~\bibnamefont{Delgado}} \bibnamefont{and}
  \bibinfo{author}{\bibfnamefont{M.}~\bibnamefont{Quiros}},
  \bibinfo{journal}{Phys.Lett.} \textbf{\bibinfo{volume}{B484}},
  \bibinfo{pages}{355} (\bibinfo{year}{2000}), \eprint{hep-ph/0004124}.

\bibitem[{\citenamefont{Batra et~al.}(2004)\citenamefont{Batra, Delgado,
  Kaplan, and Tait}}]{Batra:2003nj}
\bibinfo{author}{\bibfnamefont{P.}~\bibnamefont{Batra}},
  \bibinfo{author}{\bibfnamefont{A.}~\bibnamefont{Delgado}},
  \bibinfo{author}{\bibfnamefont{D.~E.} \bibnamefont{Kaplan}},
  \bibnamefont{and} \bibinfo{author}{\bibfnamefont{T.~M.} \bibnamefont{Tait}},
  \bibinfo{journal}{JHEP} \textbf{\bibinfo{volume}{0402}}, \bibinfo{pages}{043}
  (\bibinfo{year}{2004}), \eprint{hep-ph/0309149}.

\bibitem[{\citenamefont{Maloney et~al.}(2006)\citenamefont{Maloney, Pierce, and
  Wacker}}]{Maloney:2004rc}
\bibinfo{author}{\bibfnamefont{A.}~\bibnamefont{Maloney}},
  \bibinfo{author}{\bibfnamefont{A.}~\bibnamefont{Pierce}}, \bibnamefont{and}
  \bibinfo{author}{\bibfnamefont{J.~G.} \bibnamefont{Wacker}},
  \bibinfo{journal}{JHEP} \textbf{\bibinfo{volume}{0606}}, \bibinfo{pages}{034}
  (\bibinfo{year}{2006}), \eprint{hep-ph/0409127}.

\bibitem[{\citenamefont{Cheung and Roberts}(2013)}]{Cheung:2012zq}
\bibinfo{author}{\bibfnamefont{C.}~\bibnamefont{Cheung}} \bibnamefont{and}
  \bibinfo{author}{\bibfnamefont{H.~L.} \bibnamefont{Roberts}},
  \bibinfo{journal}{JHEP} \textbf{\bibinfo{volume}{1312}}, \bibinfo{pages}{018}
  (\bibinfo{year}{2013}), \eprint{1207.0234}.

\bibitem[{\citenamefont{Dimopoulos et~al.}(In
  preparation{\natexlab{b}})\citenamefont{Dimopoulos, Garcia~Garcia, Howe, and
  March-Russell}}]{flavorpaper}
\bibinfo{author}{\bibfnamefont{S.}~\bibnamefont{Dimopoulos}},
  \bibinfo{author}{\bibfnamefont{I.}~\bibnamefont{Garcia~Garcia}},
  \bibinfo{author}{\bibfnamefont{K.}~\bibnamefont{Howe}}, \bibnamefont{and}
  \bibinfo{author}{\bibfnamefont{J.}~\bibnamefont{March-Russell}}
  (\bibinfo{year}{In preparation}{\natexlab{b}}).

\bibitem[{\citenamefont{Ade et~al.}(2014)}]{Ade:2014xna}
\bibinfo{author}{\bibfnamefont{P.}~\bibnamefont{Ade}} \bibnamefont{et~al.}
  (\bibinfo{collaboration}{BICEP2 Collaboration}) (\bibinfo{year}{2014}),
  \eprint{1403.3985}.

\bibitem[{\citenamefont{Arkani-Hamed et~al.}(2000)\citenamefont{Arkani-Hamed,
  Dimopoulos, Kaloper, and March-Russell}}]{ArkaniHamed:1999gq}
\bibinfo{author}{\bibfnamefont{N.}~\bibnamefont{Arkani-Hamed}},
  \bibinfo{author}{\bibfnamefont{S.}~\bibnamefont{Dimopoulos}},
  \bibinfo{author}{\bibfnamefont{N.}~\bibnamefont{Kaloper}}, \bibnamefont{and}
  \bibinfo{author}{\bibfnamefont{J.}~\bibnamefont{March-Russell}},
  \bibinfo{journal}{Nucl.Phys.} \textbf{\bibinfo{volume}{B567}},
  \bibinfo{pages}{189} (\bibinfo{year}{2000}), \eprint{hep-ph/9903224}.

\bibitem[{\citenamefont{Cheung et~al.}(2011)\citenamefont{Cheung, D'Eramo, and
  Thaler}}]{Cheung:2011jq}
\bibinfo{author}{\bibfnamefont{C.}~\bibnamefont{Cheung}},
  \bibinfo{author}{\bibfnamefont{F.}~\bibnamefont{D'Eramo}}, \bibnamefont{and}
  \bibinfo{author}{\bibfnamefont{J.}~\bibnamefont{Thaler}},
  \bibinfo{journal}{JHEP} \textbf{\bibinfo{volume}{1108}}, \bibinfo{pages}{115}
  (\bibinfo{year}{2011}), \eprint{1104.2600}.

\bibitem[{\citenamefont{Aad et~al.}(2013)}]{Aad:2013ija}
\bibinfo{author}{\bibfnamefont{G.}~\bibnamefont{Aad}} \bibnamefont{et~al.}
  (\bibinfo{collaboration}{ATLAS}), \bibinfo{journal}{JHEP}
  \textbf{\bibinfo{volume}{1310}}, \bibinfo{pages}{189} (\bibinfo{year}{2013}),
  \eprint{1308.2631}.

\bibitem[{\citenamefont{Chatrchyan
  et~al.}(2013{\natexlab{a}})}]{Chatrchyan:2013xna}
\bibinfo{author}{\bibfnamefont{S.}~\bibnamefont{Chatrchyan}}
  \bibnamefont{et~al.} (\bibinfo{collaboration}{CMS Collaboration}),
  \bibinfo{journal}{Eur.Phys.J.} \textbf{\bibinfo{volume}{C73}},
  \bibinfo{pages}{2677} (\bibinfo{year}{2013}{\natexlab{a}}),
  \eprint{1308.1586}.

\bibitem[{ATL(2013{\natexlab{a}})}]{ATLAS-CONF-2013-017}
\bibinfo{type}{Tech. Rep.} \bibinfo{number}{ATLAS-CONF-2013-017},
  \bibinfo{institution}{CERN}, \bibinfo{address}{Geneva}
  (\bibinfo{year}{2013}{\natexlab{a}}).

\bibitem[{\citenamefont{Chatrchyan
  et~al.}(2013{\natexlab{b}})}]{Chatrchyan:2012oaa}
\bibinfo{author}{\bibfnamefont{S.}~\bibnamefont{Chatrchyan}}
  \bibnamefont{et~al.} (\bibinfo{collaboration}{CMS Collaboration}),
  \bibinfo{journal}{Phys.Lett.} \textbf{\bibinfo{volume}{B720}},
  \bibinfo{pages}{63} (\bibinfo{year}{2013}{\natexlab{b}}), \eprint{1212.6175}.

\bibitem[{ATL(2013{\natexlab{b}})}]{ATL-PHYS-PUB-2013-011}
\bibinfo{type}{Tech. Rep.} \bibinfo{number}{ATL-PHYS-PUB-2013-011},
  \bibinfo{institution}{CERN}, \bibinfo{address}{Geneva}
  (\bibinfo{year}{2013}{\natexlab{b}}).

\bibitem[{ATL(2013{\natexlab{c}})}]{ATLAS-CONF-2013-061}
\bibinfo{type}{Tech. Rep.} \bibinfo{number}{ATLAS-CONF-2013-061},
  \bibinfo{institution}{CERN}, \bibinfo{address}{Geneva}
  (\bibinfo{year}{2013}{\natexlab{c}}).

\bibitem[{\citenamefont{Chatrchyan
  et~al.}(2013{\natexlab{c}})}]{Chatrchyan:2013iqa}
\bibinfo{author}{\bibfnamefont{S.}~\bibnamefont{Chatrchyan}}
  \bibnamefont{et~al.} (\bibinfo{collaboration}{CMS Collaboration})
  (\bibinfo{year}{2013}{\natexlab{c}}), \eprint{1311.4937}.

\end{thebibliography}

\end{document}